\documentclass[10pt,letterpaper,twocolumn,english,aps,prl,showpacs,superscriptaddress,floatfix,longbibliography]{revtex4-2}
\usepackage{mathptmx}
\usepackage[T1]{fontenc}
\usepackage[utf8]{inputenc}
\usepackage{xcolor}
\usepackage{babel}
\usepackage{bm}
\usepackage{amsmath}
\usepackage{amssymb}
\usepackage{cancel}
\usepackage{graphicx}
\usepackage[pdfusetitle,
 bookmarks=true,bookmarksnumbered=false,bookmarksopen=false,
 breaklinks=false,pdfborder={0 0 1},backref=false,colorlinks=true]
 {hyperref}
\hypersetup{
 linkcolor=blue,anchorcolor=red,citecolor=blue,urlcolor=blue}

\makeatletter

\pdfpageheight\paperheight
\pdfpagewidth\paperwidth

\usepackage{babel}
\usepackage{calrsfs}
\DeclareMathAlphabet{\pazocal}{OMS}{zplm}{m}{n}
\usepackage{amsfonts}
\usepackage{mathrsfs}
\usepackage{bm}
\usepackage{xcolor}
\usepackage{braket}
\usepackage{subcaption}

\newlength{\seplinewidth}
\newlength{\seplinesep}
\colorlet{sepline}{orange}

\usepackage{ragged2e}

\DeclareCaptionLabelSeparator{periodspace}{.~}
\addto\captionsenglish{\renewcommand{\figurename}{FIG.}}
\renewcommand{\thefigure}{\arabic{figure}}

\def\fnum@figure{\figurename~\thefigure.}

\AtBeginDocument{\DeclareCaptionJustification{justified}{\justifying}\captionsetup{justification=justified,singlelinecheck=false}}
\makeatother

\begin{document}
\title{Magnetic Quadrupole Lens and Current-Induced Quadrupolar Dynamics
of Antiskyrmion}
\author{Linfeng Du}
\thanks{These authors contributed equally to this work.}
\affiliation{School of Physics and State Key Laboratory of Electronic Thin Films
and Integrated Devices, University of Electronic Science and Technology
of China, Chengdu 611731, China}
\author{Yang Liu}
\thanks{These authors contributed equally to this work.}
\affiliation{School of Physics and State Key Laboratory of Electronic Thin Films
and Integrated Devices, University of Electronic Science and Technology
of China, Chengdu 611731, China}
\author{Peng Yan}
\email{yan@uestc.edu.cn}

\affiliation{School of Physics and State Key Laboratory of Electronic Thin Films
and Integrated Devices, University of Electronic Science and Technology
of China, Chengdu 611731, China}
\author{Ying Su}
\email{yingsu@uestc.edu.cn}

\affiliation{School of Physics and State Key Laboratory of Electronic Thin Films
and Integrated Devices, University of Electronic Science and Technology
of China, Chengdu 611731, China}
\begin{abstract}
We demonstrate that antiskyrmions can serve as magnetic quadrupole
lenses that focus or defocus electric currents traversing them. We
showcase the antiskyrmion core as a focal point where electron beams
converge for specific injection directions, leading to anisotropic
and nonuniform current distributions. Thus the current-induced antiskyrmion
dynamics is profoundly modified by the magnetic quadrupole lensing
via the spin-transfer torque. By incorporating quantum transport and
micromagnetic simulations iteratively, we self-consistently capture
the coupled dynamics of conduction electrons and antiskyrmions. Our
results unveil a quadrupolar antiskyrmion Hall effect that is highly
anisotropic with respect to the current direction and can be effectively
tuned by the spin-orbit coupling. In particular, the current focusing
(defocusing) dramatically enhances (suppresses) the antiskyrmion speed
while simultaneously reducing its Hall angle, thereby overcoming a
key challenge for antiskyrmion racetrack memories. Our work establishes
the intrinsic current lensing as a powerful mechanism for manipulating
topological magnetic textures and opens new avenues for antiskyrmion-based
spintronics. 
\end{abstract}
\date{\today}

\maketitle
\textcolor{blue}{\emph{Introduction}}---Just as optical lenses focus
light, magnetic multipole lenses are designed to focus charged-particle
beams through the Lorentz force and play indispensable roles in various
applications, from electron microscopy to particle accelerators \cite{orloff2017handbook}.
Owing to the wave-particle duality of electrons, their ballistic transport
in solids closely parallels light propagation in optical media, laying
the foundation for solid-state electron optics \cite{washburn1990electron,van1995principles,dragoman1999optical}.
Within this framework, electron lensing has emerged as a powerful
paradigm for steering current flow and has been realized in gated
two-dimensional (2D) electron gases \cite{Sivan1990Electrostatic,spector1990electron,topinka2001coherent}
and graphene \textit{p-n} junctions \cite{lee2015observation,chen2016electron,Liu2017Creating,Burrow2026Ballistic}.
These advances demonstrate that electron trajectories can be tailored
through engineered electrostatic potentials \cite{Raedt1989Focused,cheianov2007focusing}.
By contrast, a condensed-matter counterpart of the magnetic multipole
lens, capable of focusing electric current through intrinsic magnetic
textures, remains elusive. A central challenge is to realize magnetic
multipole fields on length scales comparable to the electron mean
free path, enabling coherent manipulation of electron flow.

Topological magnetic textures, such as skyrmions, have attracted considerable
attention as robust information carriers in spintronics \cite{fert2013skyrmions,romming2013writing,sampaio2013nucleation,jiang2015blowing,kovalev2018skyrmions,Petrovi2025Quantum,Zhou2025Topological}.
When spin-polarized particles traverse these swirling spin configurations
adiabatically, they experience emergent magnetic fields that deflect
their trajectories through the Lorentz force \cite{volovik1987linear,Ye1999Berry,Bruno2004Topological,Nagaosa2012Emergent,nagaosa2013topological,van2013Magnetic,Lan2021Skew},
leading to the topological Hall effect \cite{Lee2009Unusual,Neubauer2009Topological,Kanazawa2011Large},
magnonic Hall effect \cite{mochizuki2014thermally,Weber2022Topological},
etc. Based on this mechanism, magnetic hopfions and antiferromagnetic
skyrmions have recently emerged as magnonic lenses that focus incoming
spin waves \cite{Saji2023Hopfion,Wu2026Antiferromagnetic}. In these
systems, the emergent magnetic fields are isotropic in the radial
direction and the topological magnetic textures are treated as static.
On the other hand, antiskyrmions with broken rotational symmetry and
quadrupole magnetic charges \cite{nayak2017magnetic,karube2021room,He2022Visualizing}
are capable to generate magnetic quadrupole fields (MQF) acting on
conduction electrons.\textcolor{red}{{} }However, it remains unclear
how conduction electrons respond to the antiskyrmion-induced MQF.
Meanwhile, the tailored electron trajectory would in turn modify the
current-induced dynamics of antiskyrmions through the spin-transfer
torque (STT) \cite{zhang2004roles}, which has yet to be explored.

\begin{figure}
\begin{centering}
\includegraphics[width=8.5cm]{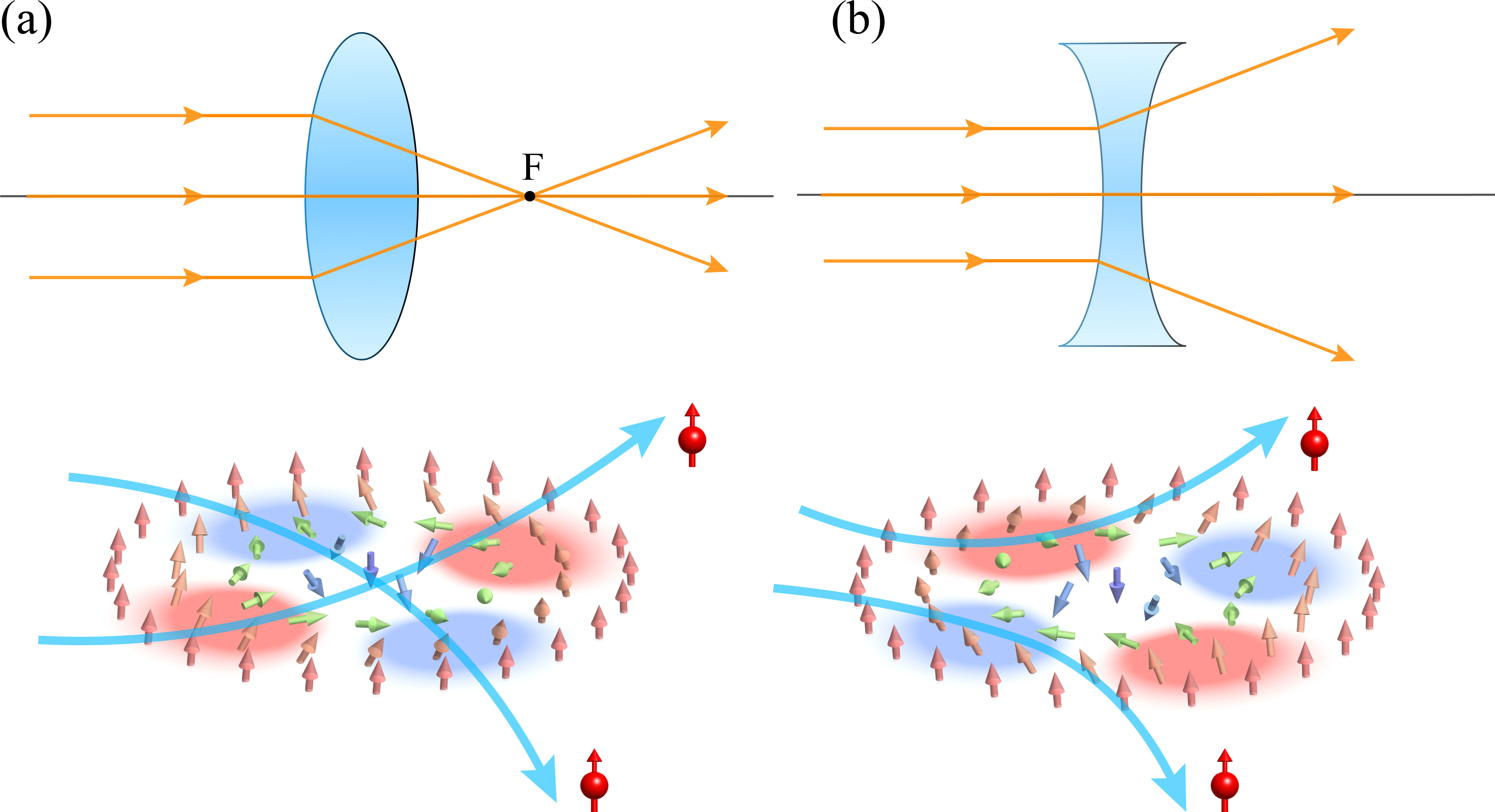} 
\par\end{centering}
\caption{Analogy between optical lenses (upper panels) and antiskyrmion magnetic
quadrupole lenses (lower panels). Focusing optical and magnetic quadrupole
lenses are shown in (a), while their defocusing counterparts are shown
in (b). The emergent magnetic quadrupole field of the antiskyrmion
is represented by the alternating blue and red shading regions. The
incident electrons are focused at the antiskyrmion core, which serves
as the focal point of the magnetic quadrupole lens in (a).}
\label{fig:1} 
\end{figure}

In this paper, we address these questions by investigating the coupled
dynamics of conduction electrons and antiskyrmions under the influence
of MQF and STT arising from each other, respectively. Employing the
iterative quantum transport and micromagnetic simulations, we self-consistently
capture the current distribution and antiskyrmion trajectory. We demonstrate
that antiskyrmions can function as magnetic quadrupole lenses, focusing
or defocusing conduction electrons injected along two orthogonal directions,
as shown in Fig. \ref{fig:1}. The antiskyrmion core serves as the
focal point of the magnetic quadrupole lens, where the focused electron
beams converge. Therefore, the electric current traversing the antiskyrmion
becomes highly anisotropic and nonuniform, profoundly modifying the
current-induced antiskyrmion dynamics. Indeed, our results unveil
a quadrupolar antiskyrmion Hall effect (QASkHE) that exhibits a pronounced
quadrupolar dependence on the current direction and can be effectively
tuned by the Rashba spin-orbit coupling (RSOC). In particular, the
antiskyrmion velocity is substantially enhanced (suppressed) by the
focused (defocused) current through its core. Meanwhile, the corresponding
Hall angle can be reduced to zero, eliminating the transverse motion
that limits practical applications such as racetrack memories \cite{parkin2008magnetic,fert2013skyrmions,Zhou2025Topological}.
To elucidate the underlying mechanism, we employ a generalized Thiele
equation that incorporates the spatially nonuniform current distribution
and quantitatively captures the QASkHE. Our work thus establishes
a new paradigm for manipulating the electron transport and antiskyrmion
dynamics through magnetic quadrupole lensing.

\begin{figure}
\begin{centering}
\includegraphics[width=8.5cm]{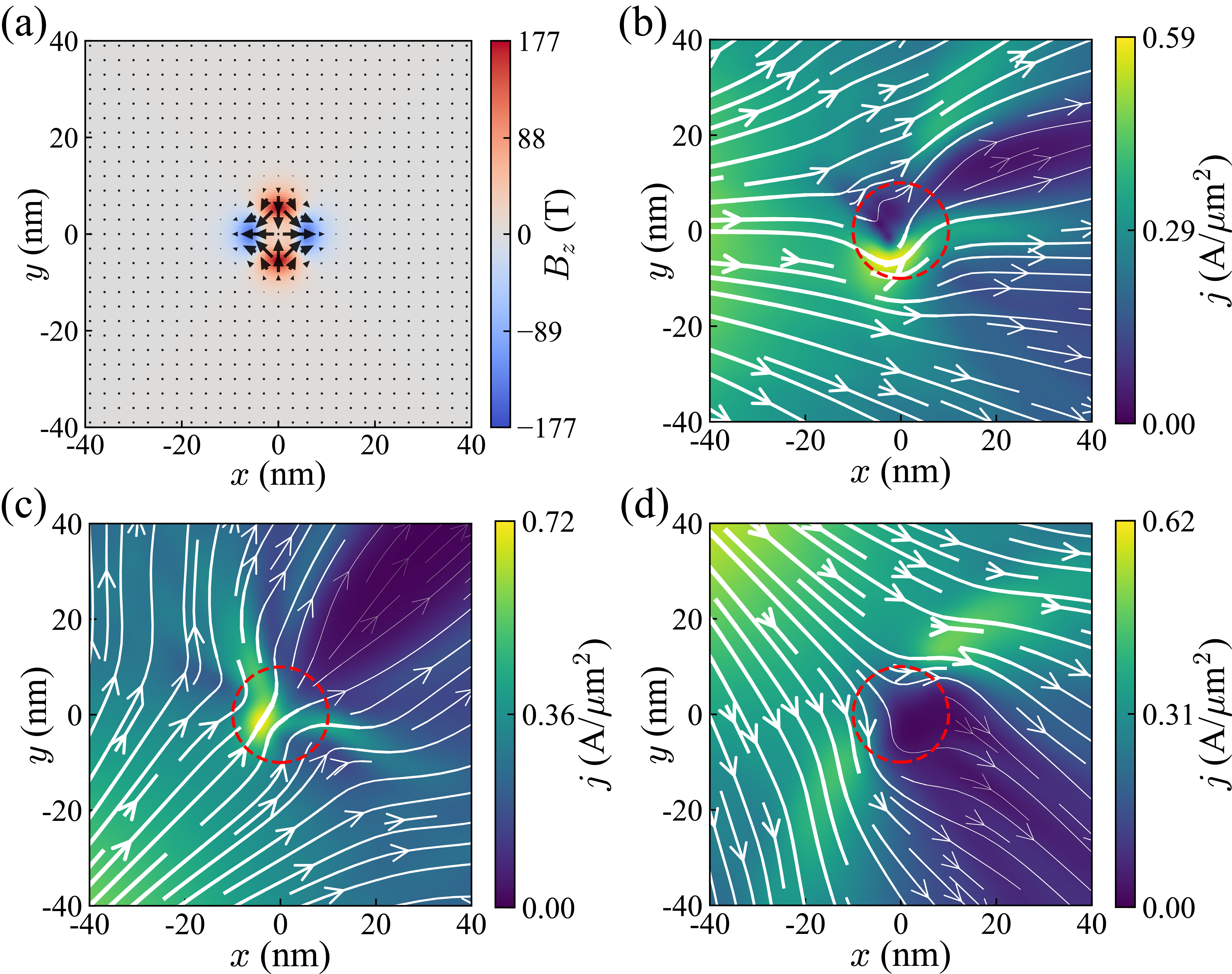}
\par\end{centering}
\caption{(a) Emergent magnetic field of an antiskyrmion at finite RSOC. (b)-(d)
Spatial distributions of electron currents injected at $\theta_{\mathrm{in}}=0^{\circ}$,
$45^{\circ}$, and $-45^{\circ}$, respectively. The white streamlines
indicate the local electron current direction. The red dashed circles
in (b)-(d) indicate the position of the antiskyrmion shown in (a).
For $\theta_{\mathrm{in}}=\pm45^{\circ}$, the electron current is
focused and defocused at the antiskyrmion core, as respectively shown
in (c) and (d).}
\label{fig:2}
\end{figure}

\textcolor{blue}{\emph{Model}}---We consider an antiskyrmion stabilized
by the anisotropic Dzyaloshinskii--Moriya interaction (DMI) \cite{huang2017stabilization,hoffmann2017antiskyrmions,Camosi2018Micromagnetics},
as described by the classical Heisenberg model

\begin{equation}
\begin{aligned}\mathcal{H}_{S}= & -A\sum_{\langle\bm{r},\bm{r}^{\prime}\rangle}\bm{S}_{\bm{r}}\cdot\bm{S}_{\bm{r}^{\prime}}-K\sum_{\bm{r}}(S^{z}_{\bm{r}})^{2}\\
 & +D\sum_{\bm{r}}\left[(\bm{S}_{\bm{r}}\times\bm{S}_{\bm{r}+\hat{\bm{x}}})\cdot\hat{\bm{y}}+(\bm{S}_{\bm{r}}\times\bm{S}_{\bm{r}+\hat{\bm{y}}})\cdot\hat{\bm{x}}\right]
\end{aligned}
\label{eq:Hs}
\end{equation}
on a square lattice. Here the first two terms are the nearest-neighbor
exchange interaction and out-of-plane easy-axis anisotropy, respectively.
The last term encodes the anisotropic DMI, which favors spin spirals
with opposite chiralities along two orthogonal axes \cite{he2024experimental}.
To be concrete, we set $A=6~\mathrm{meV}$, $D=0.6~\mathrm{meV}$,
and $K=0.11~\mathrm{meV}$ in Eq. (\ref{eq:Hs}). The competition
among these interactions stabilizes the antiskyrmion shown in Fig.
\ref{fig:2}(a). This unique topological spin texture can be parameterized
as 
\begin{equation}
\bm{S}(\bm{r})=\left[\cos(n\phi+{\color{red}{\normalcolor \eta}})\sin\theta(r),\sin(n\phi+{\color{red}{\normalcolor \eta}})\sin\theta(r),\cos\theta(r)\right],
\end{equation}
where $(r,\phi)$ are the polar coordinates in real space. Here $n\phi+\eta$
and $\theta(r)$ denote the azimuthal and polar angles of the local
magnetic moment, with ${\normalcolor n}$ and ${\normalcolor \eta}$
specifying the vorticity and helicity of the spin configuration \cite{nagaosa2013topological}.
Note that skyrmion and antiskyrmion have opposite vorticities $n=\pm1$,
while $\eta$ determines the orientation (Bloch or Néel type) of the
antiskyrmion (skyrmion). For the antiskyrmion in Fig. \ref{fig:2}(a),
it has $n=-1$, $\eta=0$, and boundary conditions of $\theta(0)=\pi$
and $\theta(\infty)=0$.

A conduction electron traversing the topological magnetic texture
can be described by the Hamiltonian 
\begin{equation}
\mathcal{H}=\frac{\bm{p}^{2}}{2m^{*}}+\frac{\lambda}{\hslash}\left(\bm{\sigma}\times\bm{p}\right)\cdot\hat{\bm{z}}-J\bm{\sigma}\cdot\bm{S}(\bm{r}),\label{eq:Hamiltonian}
\end{equation}
where the first term is the kinetic energy with the effective mass
$m^{*}$, the second term represents the RSOC, and the last term encodes
the exchange coupling between electron spin $\bm{\sigma}$ with localized
spin $\bm{S}(\bm{r})$. In the strong-exchange limit, the electron
spin adiabatically follows the smooth magnetic texture during propagation,
giving rise to the emergent magnetic field acting on conduction electrons
\cite{volovik1987linear,Ye1999Berry,Bruno2004Topological,Nagaosa2012Emergent,nagaosa2013topological}.\textcolor{red}{{}
}This field can be derived by applying a spatially dependent unitary
transformation to the Hamiltonian Eq. (\ref{eq:Hamiltonian}), which
aligns the spin quantization axis with local magnetic moments. Upon
projecting onto the partially occupied spin-up sector, one obtains
the emergent magnetic fields 
\begin{flalign}
B^{T}_{z} & =\frac{\hslash}{2e}\bm{S}\cdot\left(\frac{\partial\bm{S}}{\partial x}\times\frac{\partial\bm{S}}{\partial y}\right)=\frac{\hslash}{2e}\frac{n}{r}\sin\theta\frac{d\theta}{dr},\label{eq:B_J}
\end{flalign}
\begin{equation}
\begin{aligned}B^{M}_{z} & =\frac{m^{*}\lambda}{e\hslash}\nabla\cdot\bm{S}\\
 & =\frac{m^{*}\lambda}{e\hslash}\cos[(n-1)\phi+{\color{red}{\normalcolor \eta}}]\left(\cos\theta\frac{d\theta}{dr}+\frac{n}{r}\sin\theta\right),
\end{aligned}
\label{eq:B_lambda}
\end{equation}
due to the interplay among the exchange coupling, RSOC, and topology
of the spin configuration (see Supplemental Material \cite{supplement}
for details). Here $B^{T}_{z}$ and $B^{M}_{z}$ arise from the topological
charge density $\rho_{T}=\frac{1}{4\pi}\bm{S}\cdot\left(\partial_{x}\bm{S}\times\partial_{y}\bm{S}\right)$
and magnetic charge density $\rho_{M}=\nabla\cdot\bm{S}$, respectively
\cite{Kim2013Chirality}. In contrast to $B^{T}_{z}$, which is determined
solely by the spin texture, $B^{M}_{z}$ depends also on the effective
mass $m^{*}$ and RSOC strength $\lambda$, providing an additional
knob for tuning the emergent magnetic field.

\textcolor{blue}{\emph{Antiskyrmion magnetic quadrupole lens}}---The
topological Hall effect is conventionally attributed to the Lorentz
force arising from the emergent magnetic field $B^{T}_{z}\propto\rho_{T}$
of topological magnetic textures. Despite their broken rotational
symmetry, antiskyrmions exhibit an isotropic topological Hall effect
because $B^{T}_{z}$ is independent of the polar angle $\phi$ in
Eq. (\ref{eq:B_J}). However, the fourfold rotoinversion symmetry
of antiskyrmions gives rise to the quadrupole magnetic charge \cite{nayak2017magnetic,karube2021room,He2022Visualizing},
in contrast to the monopole magnetic charge of skyrmions. Therefore,
the RSOC-induced $B^{M}_{z}\propto\rho_{M}$ manifests as a magnetic
quadrupole (monopole) field of the antiskyrmion (skyrmion), which
is encoded in the polar-angle-dependent factor $\cos[(n-1)\phi+\eta]$
with $n=-1$ $(+1)$ in Eq. (\ref{eq:B_lambda}). Indeed, an anisotropic
topological Hall effect has been observed in the antiskyrmion-hosting
tetragonal compound Mn$_{1.4}$PtSn \cite{Vir2019Anisotropic}. It
points to a potentially overlooked contribution from $B^{M}_{z}$
that inherits the fourfold rotoinversion symmetry of antiskyrmions.
Remarkably, $B^{M}_{z}$ can dominate over $B^{T}_{z}$ for smooth
magnetic textures, as the former is linear whereas the latter is quadratic
in the spatial gradients of the magnetization. Here we consider a
moderate antiskyrmion radius $R=10\,\mathrm{nm}$, RSOC $\lambda=0.3\,\mathrm{eV\cdot nm}$,
and effective electron mass $m^{*}=0.17m_{e}$, the magnitude of the
two emergent fields can be estimated as $\left|B^{T}_{z}\right|\sim\pi^{2}\hslash/2eR^{2}=32.5\,\mathrm{T}$
and $\left|B^{M}_{z}\right|\sim\pi m^{*}\lambda/e\hslash R=138.4\,\mathrm{T}$.
Thus $B^{M}_{z}$ substantially exceeds $B^{T}_{z}$, rendering the
total emergent field $B_{z}=B^{T}_{z}+B^{M}_{z}$ predominantly quadrupolar,
as shown in Fig. \ref{fig:2}(a).

How conduction electrons respond to the emergent MQF of an antiskyrmion
remains unexplored. In order to address this question, we investigate
the quantum transport of conduction electrons traversing an antiskyrmion.
Here we consider a four-terminal Hall bar in which an antiskyrmion
is stabilized by Eq. (\ref{eq:Hs}) at the center of the scattering
region, as shown in Fig. \ref{fig:2}(a). The system size is $200\,\mathrm{nm}\times200\,\mathrm{nm}$
with the lattice constant $a=0.5\,\mathrm{nm}$. The conduction electron
hopping on the square lattice is subjected to the RSOC and exchange
coupling with localized spins, as depicted by Eq. (\ref{eq:Hamiltonian})
(see Supplemental Material \cite{supplement} for the tight-binding
approach). A uniform electron current $\bm{j}_{\mathrm{in}}=j_{\mathrm{in}}\left(\cos\theta_{\mathrm{in}},\sin\theta_{\mathrm{in}}\right)$
is injected along the direction specified by the incident angle $\theta_{\mathrm{in}}$.
The local current distribution can be obtained from the expectation
value of the bond current operator $j_{\bm{r}\bm{r}^{\prime}}=i\frac{e}{\hslash}\left(c^{\dagger}_{\bm{r}}H_{\bm{r}\bm{r}^{\prime}}c_{\bm{r}^{\prime}}-c^{\dagger}_{\bm{r}^{\prime}}H^{\dagger}_{\bm{r}\bm{r}^{\prime}}c_{\bm{r}}\right)$
where $c^{\dagger}_{\bm{r}}$ ($c_{\bm{r}}$) is the spinor creation
(annihilation) operator at site $\bm{r}$ and $H_{\bm{r}\bm{r}^{\prime}}$
denotes the hopping matrix between two nearest-neighbor sites. For
concreteness, we set $J=2\,\mathrm{eV}$ and Fermi energy $E_{F}=-5.95\,\mathrm{eV}$
for the strong-exchange limit $J\gg\lambda k_{F}$.

When conduction electrons are injected in the horizontal direction
($\theta_{\mathrm{in}}=0^{\circ}$ or $180^{\circ}$), they bypass
the antiskyrmion along semicircular trajectories through the alternating
MQF surrounding its core, as shown in Fig. \ref{fig:2}(b). Similar
behaviors have been observed for the current incident in the vertical
direction ($\theta_{\mathrm{in}}=\pm90^{\circ}$). However, for the
incoming electrons along the diagonal directions ($\theta_{\mathrm{in}}=\pm45^{\circ}$
or $\mp135^{\circ}$), the current distribution becomes qualitatively
different. In this case, the conduction electrons can be separated
into two beams with equal partitions that enter regions of opposite
emergent magnetic fields, as sketched in Fig. \ref{fig:1}. Depending
on the injection direction, the two electron beams are deflected by
opposite Lorentz forces either toward or away from each other for
$\theta_{\mathrm{in}}=\pm45^{\circ}$ (or $\mp135^{\circ}$), as respectively
shown in Fig. \ref{fig:2}(c) and \ref{fig:2}(d). Therefore, the
antiskyrmion can function as a magnetic quadrupole lens that focuses
or defocuses incoming electrons along the two orthogonal diagonal
directions. In particular, the focused electron beams converge at
the antiskyrmion core, which serves as the focal point of the magnetic
quadrupole lens, see Fig. \ref{fig:2}(c). 

\begin{figure}
\begin{centering}
\includegraphics[width=8.5cm]{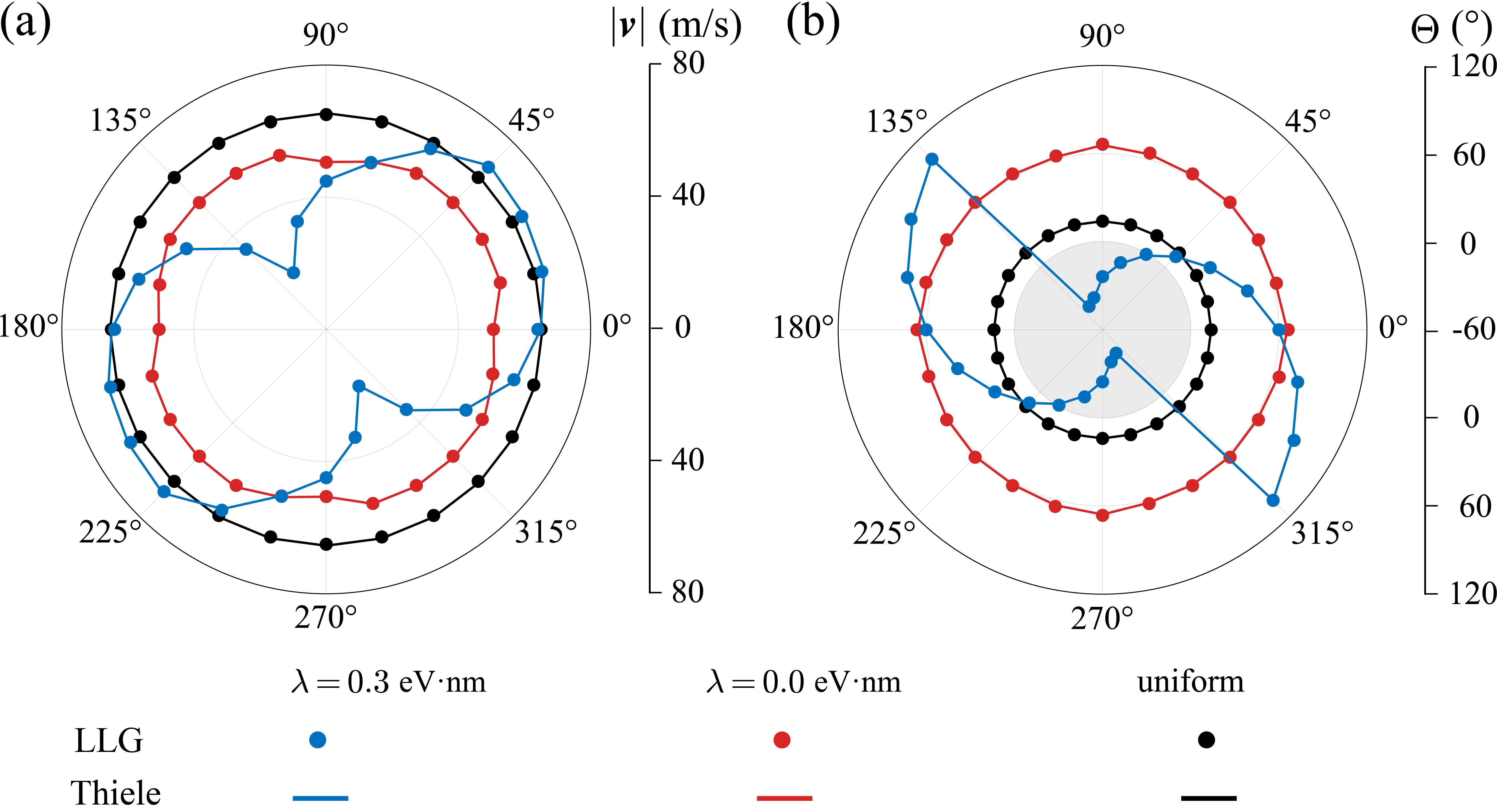} 
\par\end{centering}
\caption{(a) and (b) Polar plots of the antiskyrmion speed $v$ and Hall angle
$\Theta$ as functions of the electron current incident angle $\theta_{\mathrm{in}}$
for different $j_{\mathrm{in}}$. (c) and (d) The $v$ and $\Theta$
for finite RSOC, vanishing RSOC, and uniform-current approximation.
The self-consistent results represented by open circles are obtained
by solving the LLG Eq. (\ref{eq:llgs_stt}) with the current $\bm{j}(\bm{r})$
updated from the quantum transport calculations. The corresponding
results from the generalized Thiele Eq. (\ref{eq:Thiele}) are represented
by lines of the same color as the open circles.}
\label{fig:3} 
\end{figure}

\textcolor{blue}{\emph{Current-induced quadrupolar antiskyrmion dynamics}}---So
far, we have assumed that the antiskyrmion remains static in the quantum
transport calculations. However, a spin-polarized current transfers
angular momentum to the local moments, generating STT that drives
antiskyrmion dynamics. We now allow the antiskyrmion to evolve dynamically
according to the Landau-Lifshitz-Gilbert (LLG) equation
\begin{equation}
\begin{aligned}\frac{\partial\bm{S}}{\partial t}= & -\gamma\bm{S}\times H_{\mathrm{eff}}+\alpha\bm{S}\times\frac{\partial\bm{S}}{\partial t}+\frac{Pa^{3}}{e\left(1+\beta^{2}\right)}\left[\bigl(\bm{j}(\bm{r})\cdot\nabla\bigr)\bm{S}\right]\\
 & -\frac{\beta Pa^{3}}{e\left(1+\beta^{2}\right)}\bm{S}\times\left[\bigl(\bm{j}(\bm{r})\cdot\nabla\bigr)\bm{S}\right],
\end{aligned}
\label{eq:llgs_stt}
\end{equation}
where $\gamma=g\mu_{B}/\hslash$ is the gyromagnetic ratio, $\alpha$
is the Gilbert damping constant, and $\bm{H}_{\mathrm{eff}}=-\frac{1}{\hslash\gamma}\frac{\delta\mathcal{H}_{S}}{\delta\bm{S}(\bm{r})}$\textcolor{black}{{}
is the effective field. }The third and fourth terms describe the adiabatic
and nonadiabatic STTs, respectively \cite{zhang2004roles}. Here $P$
is the spin polarization and $\beta$ is the nonadiabaticity parameter.
In the strong-exchange limit, the electron spin remains adiabatically
aligned with the local magnetic moment, with RSOC-induced spin mixing
suppressed by the small parameter $\lambda k_{F}/J\ll1$. The Rashba
spin-orbit torque, which arises from the misalignment between electron
spin and local magnetization \cite{Manchon2008Theory}, is therefore
subleading to the exchange-mediated STT and is neglected here.

To investigate how magnetic quadrupole lensing affects antiskyrmion
dynamics, we first obtain the current density $\bm{j}(\bm{r})$ from
quantum transport calculations and use it to evolve the spin texture
$\bm{S}(\bm{r})$ according to Eq. (\ref{eq:llgs_stt}). The updated
spin texture is then fed back into the quantum transport calculation
at the next time step. By iterating this procedure, we self-consistently
capture the coupled dynamics of the conduction electron and antiskyrmion
(see Supplemental Movies). The antiskyrmion velocity $\bm{v}$ can
be obtained by tracing the motion of its center of mass. The dimensionless
parameters in Eq. (\ref{eq:llgs_stt}) are fixed at $\alpha=0.2$,
$P=0.2$, and $\beta=0.01$. For different incident electron currents
$\bm{j}_{\mathrm{in}}$, the antiskyrmion speed $v=\left|\bm{v}\right|$
and Hall angle $\Theta=\arg(\bm{v})-\theta_{\mathrm{in}}$ are respectively
shown in Figs. \ref{fig:3}(a) and \ref{fig:3}(b). Apparently, both
$v$ and $\Theta$ (open circles) exhibit a pronounced quadrupolar
dependence on $\theta_{\mathrm{in}}$, leading to the QASkHE at finite
RSOC. The antiskyrmion speed reaches its maximum $v_{\mathrm{max}}$
(minimum $v_{\mathrm{min}}$) around $\theta_{\mathrm{in}}=45^{\circ}$
and $-135^{\circ}$ ($-45^{\circ}$ and $135^{\circ}$), corresponding
to the current focusing (defocusing) directions. Meanwhile, the antiskyrmion
Hall angle $\Theta$ vanishes around these diagonal directions, which
can be utilized to eliminate its transverse motion. As the incident
current ramps up, $v$ scales linearly with $j_{\mathrm{in}}$, while
$\Theta$ remains unchanged. The highly anisotropic QASkHE with $v_{\mathrm{max}}/v_{\mathrm{min}}\approx3.5$
arises from the magnetic quadrupole lensing, enabling directional
control of both the antiskyrmion speed and Hall angle. 

For comparison, we also show the results for finite RSOC (green circles),
vanishing RSOC (brown circles), and uniform-current approximation
(black circles) in Figs. \ref{fig:3}(c) and \ref{fig:3}(d). Here
we fix the incident current with $j_{\mathrm{in}}=0.43\,\mathrm{A/\mu m^{2}}$.
In the absence of RSOC ($\lambda=0$ and $B^{M}_{z}=0$), the emergent
magnetic field reduces to the isotropic $B_{z}=B^{T}_{z}$, such that
$v\approx53.4\pm2.0\,\mathrm{m/s}$ and $\Theta\approx63.5\pm1.9^{\circ}$
are both nearly independent of $\theta_{\mathrm{in}}$. The small
residual variations in $v$ and $\Theta$ originate from the $\mathcal{C}_{4}$
symmetry of the underlying square lattice, rather than from the QASkHE.
On the other hand, the electric current $\bm{j}(\bm{r})$ in the LLG
Eq. (\ref{eq:llgs_stt}) is usually assumed to be uniform. Under this
approximation, the scattering of electron current by the antiskyrmion
is neglected, and the LLG equation is solved directly with $\bm{j}(\bm{r})=-\bm{j}_{\mathrm{in}}$.
Compared with the self-consistent results, it yields larger $v=65.0\,\mathrm{m/s}$
and smaller $\Theta=13.9^{\circ}$, indicating that the scattered
electron current significantly influences the antiskyrmion dynamics. 

\begin{figure}
\begin{centering}
\includegraphics[width=8.5cm]{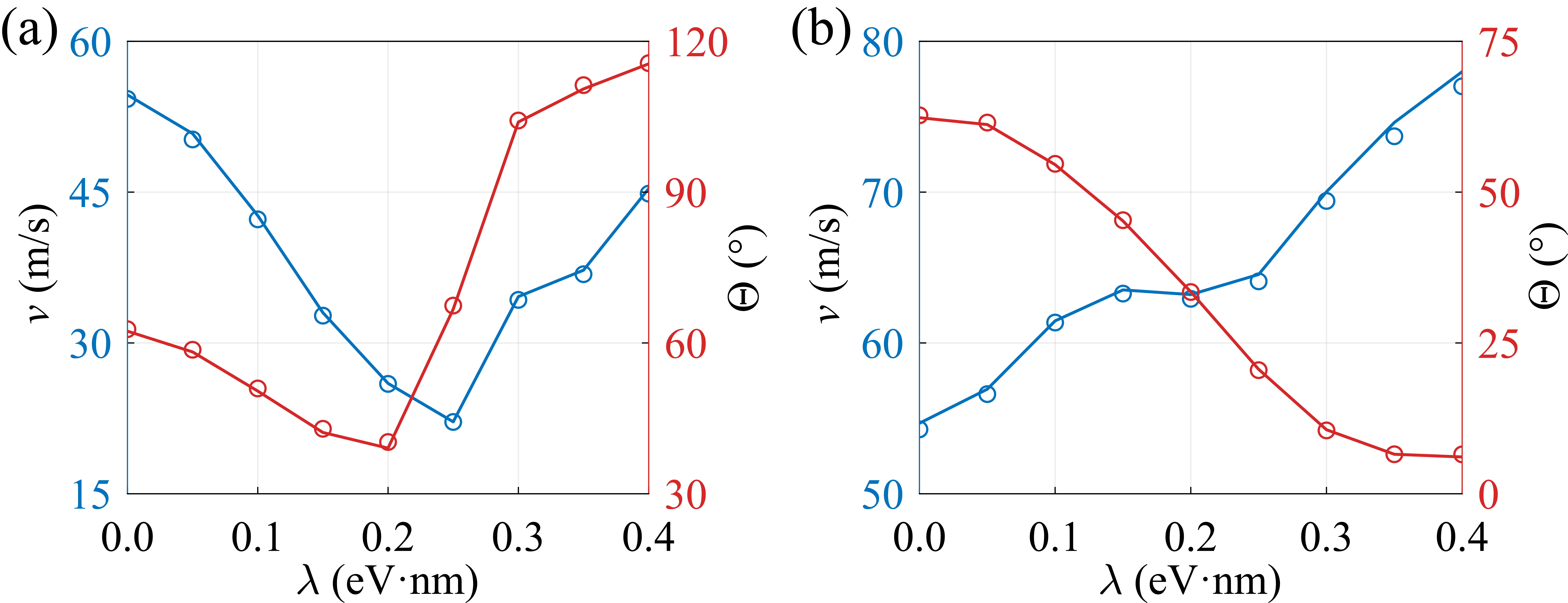} 
\par\end{centering}
\caption{(a) and (b) The antiskyrmion speed $v$ (blue) and Hall angle $\Theta$
(red) as functions of the RSOC strength $\lambda$ for electron currents
injected at $\theta_{\mathrm{in}}=-45^{\circ}$ and $45^{\circ}$,
respectively. The open circles and lines represent numerical results
obtained by solving the LLG Eq. (\ref{eq:llgs_stt}) and Thiele Eq.
(\ref{eq:Thiele}), respectively. In both equations, the current density
$\bm{j}(\bm{r})$ is obtained from the quantum transport calculations.}
\label{fig:4} 
\end{figure}

Since the antiskyrmion MQF $B^{M}_{z}$ depends linearly on $\lambda$,
the ratio between $B^{M}_{z}$ and $B^{T}_{z}$ can be continuously
tuned by the RSOC. To elucidate how the QASkHE emerges from its isotropic
counterpart as RSOC ramps up, we show the antiskyrmion speed $v$
and Hall angle $\Theta$ as a function of $\lambda$ in Fig. \ref{fig:4}.
For $\theta_{\mathrm{in}}=-45^{\circ}$, there is a pronounced transition
around $\lambda_{c}=0.25\ \mathrm{eV\cdot nm}$, above (below) which
$v$ increases (decreases) with $\lambda$, as shown in Fig. \ref{fig:4}(a).
Moreover, $\Theta$ jumps above $90^{\circ}$ for $\lambda>\lambda_{c}$.
For $\theta_{\mathrm{in}}=45^{\circ}$, $v$ increases more rapidly
with $\lambda$ while $\Theta$ begins to saturate for $\lambda>\lambda_{c}$,
as shown in Fig. \ref{fig:4}(b). These results demonstrate a crossover
from the $B^{T}_{z}$-dominant regime ($\lambda<\lambda_{c}$) to
the $B^{M}_{z}$-dominant regime ($\lambda>\lambda_{c}$) as the RSOC
increases.

To gain further insights into the QASkHE, we employ a generalized
Thiele equation to describe the motion of magnetic textures driven
by nonuniform currents. If the antiskyrmion propagates as a rigid
body, its center-of-mass velocity obeys 
\begin{equation}
-\bm{G}\times\bm{v}-\alpha\bm{D}\bm{v}+\bm{F}=0,\label{eq:Thiele}
\end{equation}
where the driving force is 
\begin{equation}
\begin{aligned}F_{i}= & -\frac{Pa^{3}}{e(1+\beta^{2})}\int d^{2}\bm{r}\Bigl\{\bm{S}\times[(\bm{j}(\bm{r})\cdot\nabla)\bm{S}]\\
 & +\beta[(\bm{j}(\bm{r})\cdot\nabla)\bm{S}]\Bigr\}\cdot\partial_{i}\bm{S},
\end{aligned}
\label{eq:force}
\end{equation}
see Supplemental Material \cite{supplement} for details. In the Thiele
Eq. (\ref{eq:Thiele}), the first two terms are the Magnus force and
dissipative force, respectively \cite{thiele1973steady}. Here $D_{ij}=\int d^{2}\bm{r}\partial_{i}\bm{S}\cdot\partial_{j}\bm{S}=D\delta_{ij}$
is the dissipative force tensor and $\bm{G}=G\hat{\bm{z}}$ is the
gyromagnetic coupling vector with $G=\int d^{2}\bm{r}\bm{S}\cdot(\partial_{x}\bm{S}\times\partial_{y}\bm{S})=4\pi Q$
determined by the topological charge $Q$ of the antiskyrmion. The
last term \textbf{$\bm{F}$} stems from the STT in Eq. (\ref{eq:llgs_stt})
and encodes the nonuniform current $\bm{j}(\bm{r})$ in the integral
of Eq. (\ref{eq:force}). Solving the Thiele Eq. (\ref{eq:Thiele}),
it yields the antiskyrmion velocity 
\begin{equation}
\bm{v}=\left(\frac{\alpha DF_{x}+GF_{y}}{\alpha^{2}D^{2}+G^{2}},\frac{\alpha DF_{y}-GF_{x}}{\alpha^{2}D^{2}+G^{2}}\right),\label{eq:velocity}
\end{equation}
which is directly proportional to the driving force and hence sensitive
to the local current distribution within the antiskyrmion.

Once the local current density $\bm{j}(\bm{r})$ is known, the antiskyrmion
dynamics can be readily determined from the generalized Thiele equation.
To validate this approach, we extract $\bm{j}(\bm{r})$ from the quantum
transport simulations and substitute it into Eqs. (\ref{eq:force})
and (\ref{eq:velocity}) to obtain the antiskyrmion velocity $\bm{v}$.
As shown in Figs. \ref{fig:3} and \ref{fig:4}, the antiskyrmion
speed $v$ and Hall angle $\Theta$ calculated from the generalized
Thiele equation (solid lines) exhibit excellent agreement with the
numerical results from the iterative quantum transport and micromagnetic
simulations (open circles). This agreement demonstrates that the impact
of magnetic quadrupole lensing on antiskyrmion dynamics is fully captured
by the driving force $\bm{F}$ induced by the nonuniform current in
the generalized Thiele equation.

\textit{\textcolor{blue}{Discussion and conclusion}}---Here we compare
the QASkHE arising from the magnetic quadrupole lensing with other
mechanisms of current-induced antiskyrmion dynamics. In addition to
STT, an anisotropic antiskyrmion Hall effect can also be driven by
spin-orbit torque (SOT) \cite{huang2017stabilization,Fei2026Multiscale}.
In this case, the electron current is injected into an adjacent heavy-metal
layer, where the spin Hall effect converts it into a transverse spin
current that exerts SOT on the antiskyrmion. Owing to the anisotropic
spin texture, the SOT-induced antiskyrmion Hall angle depends linearly
on the current incident angle, while its speed remains isotropic.
In stark contrast, both antiskyrmion speed and Hall angle exhibit
nonlinear and quadrupolar dependence on the current direction in the
QASkHE, as shown in Fig. \ref{fig:3}. To eliminate transverse motion,
previous studies have demonstrated that antiskyrmions embedded in
stripe domains can be confined into one-dimensional straight motion
driven by current pulses \cite{he2024experimental,guang2024confined}.
In our case, the antiskyrmion Hall angle vanishes for currents injected
along the diagonal directions, as shown in Fig. \ref{fig:3}(b), allowing
the antiskyrmion to move straight along the current direction without
spatial confinement.

In summary, we identify the antiskyrmion as a magnetic quadrupole
lens that focuses and defocuses electron currents injected along two
orthogonal diagonal directions, respectively. We uncover a self-consistent
feedback between antiskyrmion-induced electron lensing and current-induced
antiskyrmion dynamics, giving rise to the QASkHE. In this effect,
the antiskyrmion velocity exhibits a pronounced quadrupolar dependence
on the current direction, reflecting the fourfold rotoinversion symmetry
of the spin texture. Remarkably, the antiskyrmion Hall angle can be
reduced to zero when a diagonally incident current is focused onto
the antiskyrmion core, thereby suppressing the transverse motion.
Our work reveals magnetic quadrupole lensing as a distinct mechanism
for manipulating antiskyrmion dynamics through the self-consistent
interplay between electron transport and topological magnetic textures.

\begin{acknowledgements} \textit{\textcolor{blue}{Acknowledgments}}---This
work is supported by the National Natural Science Foundation of China
(Grants No. 12504247, No. 12374103, and No. 12434003), National Key
R\&D Program of China (Grants No. 2022YFA1402802 and No. 2025YFA1411302),
and Natural Science Foundation of Sichuan Province (Grants No. 2025ZNSFSC0866
and No. 2025NSFJQ0045). Y. S. acknowledges the support of the startup
fund and Center for HPC at the University of Electronic Science and
Technology of China.

\end{acknowledgements}

\bibliography{references}

\end{document}